\documentclass{emulateapj}
\usepackage{apjfonts}

\newcommand{\eV}{\mbox{ eV}}

\newcommand{\kel}{\mbox{ K}}
\newcommand{\mkel}{\mbox{ mK}}

\newcommand{\msun}{\mbox{ M$_\odot$}}
\newcommand{\secinv}{\mbox{ s$^{-1}$}}

\newcommand{\junits}{\mbox{ erg s$^{-1}$ cm$^{-2}$ Hz$^{-1}$ sr$^{-1}$}}
\newcommand{\coolfcn}{\mbox{ erg s$^{-1}$ cm$^{3}$}}
\newcommand{\hunits}{\mbox{ km s$^{-1}$ Mpc$^{-1}$}}
\newcommand{\bq}{\begin{equation}}
\newcommand{\eq}{\end{equation}}

\begin{document}

\title{Large-Scale Structure Shocks at Low and High Redshifts}

\author{Steven R. Furlanetto\altaffilmark{1} \& Abraham Loeb\altaffilmark{2}}

\altaffiltext{1} {Division of Physics, Mathematics, \& Astronomy;
  California Institute of Technology; Mail Code 130-33; Pasadena, CA
  91125; sfurlane@tapir.caltech.edu}

\altaffiltext{2} {Harvard-Smithsonian Center for Astrophysics, 60
Garden Street, Cambridge, MA 02138; aloeb@cfa.harvard.edu}

\begin{abstract}

Cosmological simulations show that, at the present time, a substantial
fraction of the gas in the intergalactic medium (IGM) has been
shock-heated to $T \ga 10^5 \kel$.  Here we develop an analytic model
to describe the fraction of shocked, moderately overdense gas in the
IGM.  The model is an extension of the \citet{press} description for
the mass function of halos: we assume that large-scale structure
shocks occur at a fixed overdensity during nonlinear collapse.  This
in turn allows us to compute the fraction of gas at a given redshift
that has been shock-heated to a specified temperature.  We show that,
if strong shocks occur at turnaround, our model provides a reasonable
description of the temperature distribution seen in cosmological
simulations at $z \sim 0$, although it does overestimate the
importance of weak shocks.  We then apply our model to shocks at high
redshifts.  We show that, before reionization, the thermal energy of
the IGM is dominated by large-scale structure shocks (rather than
virialized objects).  These shocks can have a variety of effects,
including stripping $\la 10\%$ of the gas from dark matter minihalos,
accelerating cosmic rays, and creating a diffuse radiation background
from inverse Compton and cooling radiation.  This radiation background
develops \emph{before} the first stars form and could have measurable
effects on molecular hydrogen formation and the spin temperature of
the 21 cm transition of neutral hydrogen.  Finally, we show that
shock-heating will also be directly detectable by redshifted 21 cm
measurements of the neutral IGM in the young universe.

\end{abstract}


\section{Introduction}

One of the major achievements of cosmology over the past two decades
has been the development of a coherent and testable paradigm for
structure formation.  In the modern picture, small density
inhomogeneities imprinted on the universe during inflation grow
continuously through gravitational instability.  These density
fluctuations still have small amplitude at the time of cosmological
recombination ($z \sim 1000$), when the cosmic microwave background
(CMB) is last scattered.  At this epoch, structure formation is
straightforward to study analytically because the fluctuations are
still linear.  Observations of the CMB have shown convincing evidence
that our theories of structure formation accurately describe the
universe at this early phase (e.g., \citealt{spergel03}).  Eventually,
the perturbation amplitude grows sufficiently large for nonlinear
effects to become important: at this point the first gravitationally
bound structures begin to form.  The structure formation problem then
passes beyond the scope of analytic approaches (except in the case of
perfectly spherical collapse; \citealt{gunn72}).  The most successful
approximation for studying the properties of gravitational instability
in the nonlinear regime is the so-called Zel'dovich approximation
\citep{zeldovich70}, which uses Lagrangian (rather than Eulerian)
perturbation theory.  The theory shows that collapse will occur first
along a single axis, creating a ``pancake'' or sheet, then along a
second axis, creating a filament, before finally converging towards a
single point.

Fully non-linear treatments of structure formation had to await the
development of numerical simulations.  These simulations confirm the
behavior predicted by the Zel'dovich approximation: at low redshifts, the
matter in the universe organizes itself into a complex network of sheets
and filaments (with galaxies and galaxy clusters at their intersections)
known as the ``cosmic web.''  This paradigm has had admirable success
reproducing the large-scale distribution of galaxies in redshift surveys
(e.g., \citealt{delapparent86}) and in describing the absorption features
of the Ly$\alpha$ forest (e.g., \citealt{cen94,hernquist96,theuns98}).

While the dark matter accretes smoothly onto the cosmic web, the
baryons do not: their infall velocities onto these structures often
exceed the local sound speed and a complex network of shocks forms.
Probably most important are the well-known virial shocks that form
when halos collapse.  However, cosmological simulations also contain
many shocks \emph{outside} of collapsed halos; these fill a fair
fraction ($\sim 10\%$) of the volume of the universe
\citep{cen99,keshet03,ryu03} and play an important role in shaping the
properties of baryons in the local universe.  Simulations predict that
$\ga 20\%$ of all baryons are in the ``warm-hot intergalactic medium''
(WHIM) with $T \sim 10^5$--$10^7 \kel$ \citep{dave01-whim}, most of
which have densities below those characteristic of virialized objects.

Unfortunately, there is little analytic understanding of the
distribution of shocks in the intergalactic medium (IGM).
\citet{cen99} argued that the characteristic temperature of the gas
can be computed by simply associating the postshock temperature with a
fluctuation collapsing over the age of the universe that has just
become nonlinear at the present epoch.  This provides a rough guide
to the expected strength of the shocks but does not predict how much
gas has been shocked or give a detailed temperature distribution.
\citet{nath} expanded this treatment using the Zel'dovich
approximation to give rough estimates of the fraction of gas in the
WHIM phase, but they were not able to predict the detailed properties
of shocks at any redshift.

There has also been little consideration of the role that similar
large-scale structure shocks play at higher redshifts.  One reason
that they are significant in the local universe is that the
characteristic shock temperature ($T_{\rm sh} \sim 10^7 \kel$; see
\citealt{dave01-whim}) is much larger than the mean temperature of the
photoionized IGM ($T_{\rm IGM} \sim 10^4 \kel$).  As redshift
increases, $T_{\rm sh}$ decreases rapidly; at $z \ga 2$, the
characteristic temperature given by \citet{cen99} is not much larger
than $T_{\rm IGM}$, so we expect strong shocks to be much less common.
However, before reionization, the typical temperature of the IGM is
much smaller and shocks can again become important (despite the much
smaller-scale perturbations that collapse).  Because shocks occur in
overdense regions, where galaxies also preferentially occur, they
obviously help to determine the environments in which most galaxies
reside.  The large-scale environment has important consequences for
feedback from (and on) collapsed objects, but this point has not been
considered in the past.  

In this paper, we consider a new analytic model for the temperature
distribution of shocked, moderately overdense gas in the IGM.  We are
guided by the fact that the distribution of collapsed objects is
well-described, to a surprisingly high precision, by the
Press-Schechter formalism \citep{press}.  The key ingredients to this
model are (1) Gaussian statistics for the density field and (2)
spherical collapse of perturbations in the nonlinear regime.
Motivated by the success of this formalism, we use the same
assumptions to model shocked gas in the IGM.  We show that our simple
model provides a reasonable fit to existing (low-redshift) simulation
data if only strong shocks are considered.  In \S \ref{formalism}, we
present our formalism for the large-scale structure shocks and use it
to describe the basic properties of shocks at both low and high
redshifts.  We then move on to consider the role that high-redshift
shocks might play.  In \S \ref{dyneffects} we describe three important
effects that the shocks can have: accelerating particles to high
energies, creating a diffuse background radiation field (principally
through cooling radiation), and stripping minihalo gas.  In \S
\ref{observe} we consider the observable consequences of high-redshift
shocks.  Finally, we summarize our main conclusions in \S \ref{disc}.

Throughout the discussion we assume a flat, $\Lambda$--dominated
cosmology, with density parameters $\Omega_m = 0.3$, $\Omega_\Lambda =
0.7$, and $\Omega_b = 0.047$ in matter, cosmological constant, and
baryons, respectively.  In the numerical calculations, we assume
$h=0.7$, where the Hubble constant is $H_0 = 100 h \hunits$.  We
take a scale-invariant primordial power spectrum ($n=1$) normalized
to an amplitude of $\sigma_8=0.9$ on 8$h^{-1}$ Mpc spheres.  These
values are consistent with the most recent measurements
\citep{spergel03}.

\section{A Model of Shocks from Large-Scale Structure}
\label{formalism}

Following \citet{cen99} and \citet{keshet03}, we associate large-scale
shocks with the collapse of nonlinear objects.  The shock velocity
$v_s$ is then approximately equal to the length scale of the
perturbation divided by the age of the universe, or
\bq 
v_s \approx H(z) R_p,
\label{eq:shockvel}
\eq 
where $R_p$ is the physical radius that the region would have had
if it expanded uniformly with the Hubble flow (i.e., $4\pi
R_p^3/3=M_s/\bar{\rho}_z$, where $M_s$ is the mass of the shocked
region and $\bar{\rho}_z$ is the mean density of the universe at
redshift $z$).  The postshock temperature $T_{\rm sh}$ is then 
\bq 
T_{\rm sh} = \eta \frac{\mu m_p}{\gamma k_B} \, v_s^2,
\label{eq:shocktemp}
\eq 
where $\mu m_p$ is the mean particle mass, $\gamma$ is the adiabatic
index of the gas, and $\eta$ is the square of the Mach number of the
shock in the downstream medium.  Given a shock velocity and the sound
speed of the IGM gas, $\eta$ is fixed in terms of $\gamma$ by the jump
conditions; however, the definition of the shock velocity in equation
(\ref{eq:shockvel}) is only approximate, and we let $\eta$ encapsulate
uncertainties in both the collapse time and the relevant length scale.
In most of the following discussion we will assume strong shocks, for
which $\eta = 5/16$ (if $\gamma=5/3$).  Note that the shock
temperature depends on both the mass of the perturbation and on
redshift.  \citet{cen99} and \citet{dave01-whim} argued that choosing
$R_p$ to be the length scale associated with the characteristic
nonlinear mass at a given epoch reproduces the typical temperature of
the WHIM gas in their simulations.

In order to compute the distribution of shocks, we make an analogy
with the usual Press-Schechter mass function \citep{press}.  The
fundamental assumptions behind this mass function are: (i) the initial
density field is Gaussian, and (ii) the threshold overdensity for a
particle to be attached to a halo is given by the dynamics of a
collapsing spherical top-hat perturbation. In this model, the fraction
of mass in collapsed halos of mass $M>M_0$ is
\bq
F(M>M_0,z) = {\rm erfc}\left[ \frac{\delta_c(z)}{\sqrt{2} \sigma(M_0)}
  \right].
\label{eq:psmass}
\eq
Here $\sigma(M_0)$ is the variance in the density power spectrum
smoothed over a mass scale $M_0$ and $\delta_c(z)=\delta_{c0}/D(z)$ is
the linearly extrapolated overdensity corresponding to the collapse of
a spherical perturbation.  $D(z)$ is the cosmological growth factor
and $\delta_{c0} \approx 1.69$ depends very weakly on the assumed
cosmology.  The mass function can be obtained by simply
differentiating equation (\ref{eq:psmass}).  This mass function
matches the distribution measured in cosmological simulations to good
accuracy, although modifications including ellipsoidal (rather than
spherical) collapse do improve the fit somewhat \citep{sheth,jenkins},
though perhaps not at the highest redshifts \citep{jang01}.

Here we make the ansatz that we can identify shocks with a different
density threshold $\delta_x$ during the collapse.  The entire
Press-Schechter formalism carries over to our situation so long as we
change the density threshold from that for collapse to the new value
and choose the smoothing scale $M_{\rm sh}$ for a specified shock
temperature $T_{\rm sh}$ through equations (\ref{eq:shockvel}) and
(\ref{eq:shocktemp}).  The resulting temperature distribution is
\begin{eqnarray}
\left[ \frac{df}{d T_{\rm sh}} \right]_{\rm naive} & = & \frac{3}{\sqrt{2 \pi}}
\, \frac{M_{\rm sh}}{T_{\rm sh}} \, \left| \frac{d \ln \sigma}{d M}
\right|_{M_{\rm sh}} \nonumber \\
& & \times 
\frac{\delta_x(z)}{\sigma(M_{\rm sh},z)} \, \exp \left[ \frac{\delta_x^2(z)}{2
    \sigma^2(M_{\rm sh},z)} \right],
\label{eq:naivedfdt}
\end{eqnarray}
where $df$ is the fraction of IGM gas (by mass) that is
shocked to a temperature in the range $(T_{\rm sh},T_{\rm sh}+dT_{\rm
  sh})$.  For reference, 
\begin{eqnarray}
M_{\rm sh} & = & 10^{13} \msun \left(
\frac{\gamma}{5/3} \, \frac{0.6}{\mu} \, \frac{5/16}{\eta} \,
\frac{T_{\rm sh}}{10^6 \kel} \right)^{3/2} \nonumber \\ 
& & \times \left[ \frac{1+z}{h(z)}
  \right]^3 \left( \frac{\Omega_m}{0.3} \right), \label{eq:mequiv}
\end{eqnarray}
where $h(z) = [\Omega_m(1+z)^3+\Omega_\Lambda]^{1/2}$.  Because the
$T_{\rm sh}$--$M_{\rm sh}$ relation is itself redshift-dependent, the
redshift dependence of $\sigma$ cannot be factored out (unlike for the
mass function).  Also, we are interested only in the fraction of gas
at a given temperature, not in the number density of such objects
(which would be more sensitive to the non-spherical nature of
collapse).  In all other respects equation (\ref{eq:naivedfdt}) is
analogous to mass function calculations.

We note that both $\eta$ and $\delta_x$ are essentially free
parameters in our formalism: in the purely spherical model, shocks do
not actually appear until complete collapse, so the density threshold
and shock velocity are ad hoc additions to the model.  However, we
will show below (see \S \ref{sims}) that a simple, physically
motivated set of choices provides a reasonable description of the
existing simulation data for strongly shocked material.  A plausible
point for the shock to first appear is turnaround [$\delta_{x} \approx
1.06/D(z)$], the point at which the perturbation breaks off from the
cosmological expansion.  In purely spherical collapse, of course, no
shocks would appear until shell-crossing occurs at virialization.
However, turnaround marks the point at which the flow begins to
converge.  It is therefore reasonable to expect asymmetries in the
collapse to begin to shock the gas at this point.  Another advantage
of associating shocks with turnaround is that the shock velocity in
equation (\ref{eq:shockvel}) exactly equals the peculiar velocity of
the collapsing gas.  Because most shocks are strong, we then assume
$\eta=5/16$ for simplicity.  We expect this choice to work best for
regions with infall velocities much larger than the IGM sound speed.
If the two are comparable, asymmetries in the collapse would have to
be extremely strong for shocks to develop before shell-crossing.  In
this regime, we might therefore expect $\delta_x$ to be closer to
$\delta_c$.

We also note that our formalism does not describe (or replace) virial
shocks around collapsing halos.  The ``large-scale structure shocks''
that we describe have much lower velocities because they occur earlier
in the infall process.  If we compare the shock temperature $T_{\rm
  sh}$ on a given mass scale to the virial temperature $T_{\rm vir}$
on the same scale, we find
\bq
\frac{T_{\rm vir}}{T_{\rm sh}} \simeq 4 \frac{(1+z)^3}{h^2(z)} \left(
\frac{\gamma}{5/3} \, \frac{5/16}{\eta} \right) \left(
\frac{\Omega_m}{0.3} \right)^{2/3} \left[ \frac{\Omega_m}{\Omega_m(z)}
\right]^{1/3},
\label{eq:tcomp}
\eq
so the virial shocks will still be relatively strong.

There is one more subtlety in computing the temperature distribution:
a particular trajectory that reaches $\delta_x(z)$ at some mass
$M_{\rm sh}$ may reach $\delta_c(z)$ at another mass $M_h<M_{\rm sh}$.
Physically, the particle could belong to a galaxy embedded inside a
large-scale filament or sheet.  We wish to exclude such gas from our
distribution because it does not truly belong to the IGM.  (It will
also be at a different temperature: the gas initially virializes and
then cools.)  Fortunately, it is straightforward to identify these
particles in the Press-Schechter model.  Given that a particle has
$(\delta_x,M_{\rm sh})$, we wish to compute the probability that it
has $(\delta_c,M_h)$.  The problem is formally identical to the
``progenitor'' problem solved by \citet{lacey}, with the solution
\begin{eqnarray}
\frac{dp}{d M_h} & = & \sqrt{\frac{2}{\pi}}
\frac{\sigma(M_h)[\delta_c(z)-\delta_x(z)]}
     {[\sigma^2(M_h)-\sigma^2(M_{\rm sh},z)]^{3/2}} \nonumber \\
& & \times \left| \frac{d \sigma}{d M} \right|_{M_h} \exp \left\{
  \frac{-[\delta_c(z)-\delta_x(z)]^2} {2[\sigma^2(M_h)-\sigma^2(M_{\rm
	sh},z)]} \right\}.
\label{eq:pbound}
\end{eqnarray}
Here $dp/dM_h$ is the probability that a particle is contained in a
halo of mass $M_h$ at $z$, given that it is part of a shock with mass
scale $M_{\rm sh}$ at $z$.  Then
\bq
\frac{df}{d T_{\rm sh}} = \left[ \frac{df}{d T_{\rm sh}} \right]_{\rm
  naive} \left[ 
1 - \int_{M_{\rm min}}^{M_{\rm sh}} dM_h \frac{dp}{d M_h} \right].
\label{eq:dfdt}
\eq
We note that this prescription fits nicely into the ``merger tree''
algorithm commonly used in ``extended Press-Schechter'' studies (e.g.,
\citealt{lacey}).  To compute the probability that a halo of mass
$M_h$ will merge into a halo of mass $M_{\rm sh}$ at a later time
$z_v$, we simply substitute $\delta_x(z) \rightarrow \delta_c(z_v)$ in
equation (\ref{eq:pbound}).  Clearly, $z_v$ is fixed by the dynamics
of collapse, $D(z_v) = (\delta_x/\delta_c) D(z)$.  But this is the
same relation as between the collapse and turnaround times in the
spherical collapse model.  We therefore find that the probability for
a halo to lie in a turnaround shock of a region with mass $M_{\rm sh}$
at $z$ is the same as for the halo to lie in that region once it has
collapsed.

The correction for bound objects would not be needed if the
large-scale shocks were able to strip gas from collapsed objects.  In
the local universe, this occurs through ram pressure stripping and is
fairly inefficient: in galaxy clusters, observations and simulations
show that the timescale is short only in dense cluster cores
\citep{quilis}, and it is likely to be much longer in the low-density
IGM.  However, if the gas in a bound object is unable to cool
radiatively and is then shocked during infall, it may be able to
escape its host halo into the IGM.  We will examine the implications
of this process in detail in \S \ref{strip}.

Note that the postshock temperature depends on the whether the gas is
ionized or neutral.  Any shock with $T_{\rm sh} > 10^4 \kel$ will
ionize gas after the shock, decreasing the actual temperature of the
postshock gas by a factor $\sim 2$.  If the gas is originally neutral,
the shock will also lose energy ionizing the gas.  For simplicity, we
assume that the universe is completely ionized for all redshifts lower
than the reionization redshift, $z<z_r$, and neutral for $z>z_r$.  The
temperatures we quote for shocks before reionization do \emph{not}
include ionization by the shock; they should thus be interpreted as
the mean thermal energy per \emph{baryon}.  The reionization redshift
also affects the minimum halo mass able to colllapse.  If the universe
is neutral and cold, we take the ``filter mass'' $M_{\rm fil}$ or
time-averaged Jeans mass, as the minimum collapse mass
\citep{gnedin-hui}.  (Because turnaround shocks occur earlier in
collapse than virialization does, we actually evaluate $M_{\rm fil}$
at the redshift corresponding to collapse of the object.)
If the universe is ionized, we take the minimum mass to have $T_{\rm
vir} > 2 \times 10^5 \kel$, below which gas infall is suppressed by
photoionization heating \citep{efstathiou92,thoul95,navarro97},
although near the time of reionization this probably overestimates the
minimum mass \citep{dijkstra03}.

\subsection{Comparison to Simulations}
\label{sims}

\begin{figure}[t]
\plotone{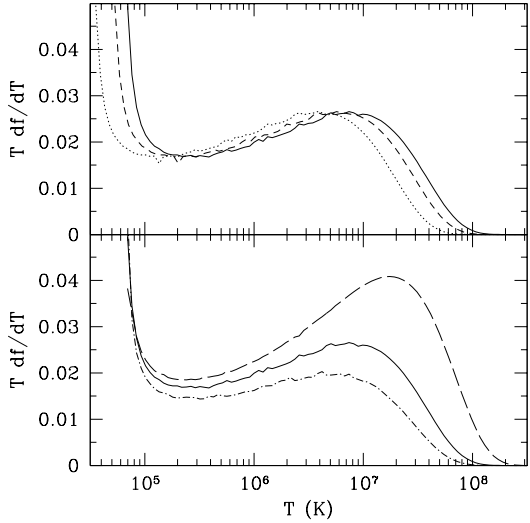}
\caption{The distribution of gas temperature at $z=0$, not including
  collapsed objects.  \emph{Top panel:} The curves have
  $\eta=0.16,\,0.24$, and $5/16$ (dotted, short-dashed, and solid lines,
  respectively); all assume $\delta_x=1.06$.  \emph{Bottom panel:} The
  curves have $\delta_x=0.8,\,1.06$, and $1.2$ (long-dashed, solid,
  and dot-dashed lines, respectively); all assume $\eta=5/16$.  }
\label{fig:z0param}
\end{figure}

While we have presented a plausible picture in which $\eta$ and
$\delta_x$ are chosen to match strong shocks at halo turnaround, these
are actually free parameters that are not fixed by our model.  In
Figure \ref{fig:z0param}, we show how their choice changes our
differential temperature distribution function, $df/dT$.  The top
panel shows $\eta=0.16,\,0.24$, and $5/16$ (dotted, short-dashed, and
solid curves, respectively), with $\delta_x=1.06$.  [Note that the
simulations of \citet{keshet03} suggest $\eta \approx 0.16$.]  The
sharp rise at low temperatures is the effect of the finite Jeans mass
that we have imposed.  Structures collapsing near the mass threshold
cannot have any gaseous subcomponents, so the fraction of gas
associated with bound objects rapidly approaches zero.  The effect of
varying $\eta$ is simply to translate the distribution along the
temperature axis: we have not changed the fraction of mass associated
with each mass scale, only the temperature associated with that scale.
The bottom panel fixes $\eta=5/16$ but sets $\delta_x=0.8,\,1.06$, and
$1.20$ (long-dashed, solid, and dash-dotted curves, respectively).
The effect here is more subtle: for example, by increasing the
threshold overdensity required for shocks, we decrease the mass
fraction at each temperature.  Because $\sigma(M)$ is not linear, the
characteristic temperature also changes (i.e., the peak of the
temperature distribution).

Comparison to Figure 5 in \citet{dave01-whim} shows that our fiducial
choice gives a reasonably good fit to the shape of the temperature
distribution at $z=0$.  We find that $\sim 13\%$ of the gas is shocked
to high temperature, compared to $\sim 25\%$ in the simulations.  The
comparison is somewhat difficult because \citet{dave01-whim} include
both bound objects (in particular galaxy groups) and gas at or near
the mean density that is nevertheless shocked to high temperatures
(see their Figure 4), while we expect our formalism to describe only
the moderately overdense gas.  We also find that our formalism
overestimates the fraction of gas with $T_{\rm sh} \sim 10^5 \kel$.
This is not surprising, because these shocks correspond to systems
fairly close to the minimum mass for galaxy formation in an ionized
medium.  The infall velocities are therefore not much larger than the
sound speed of the gas, and shocks probably develop only later in the
collapse process for such objects.  We expect our model to apply best when
the characteristic shock speed is much larger than the sound speed of
the IGM.  On the other hand, capturing weak shocks is difficult in
simulations with finite resolution \citep{keshet03,ryu03}, and they
may underestimate the fraction of gas shocked to moderate
temperatures.

\subsection{Temperature Evolution of the IGM}
\label{tempevol}

\begin{figure}[t]
\plotone{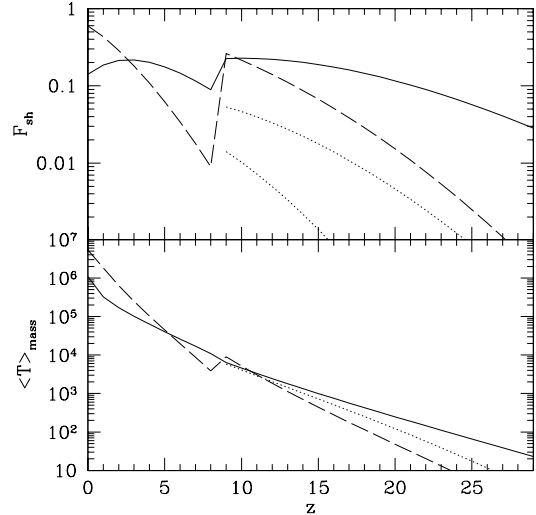}
\caption{\emph{Top panel:} The fraction of gas in the shocked phase
  (solid curve) and in collapsed objects (long-dashed curve).  The
  upper and lower dotted curves show the fraction with $T_{\rm
  sh}>10^4 \kel$ and $10^5 \kel$, respectively.  We have assumed that
  reionization occurs instantaneously at $z_r=8$, at which point the
  minimum halo mass increases dramatically. \emph{Bottom panel:} The
  mass-averaged temperature of the IGM, including shocked gas only
  (solid curve) and collapsed objects only (dashed curve).  The dotted
  curve shows the temperature including only shocks with $T_{\rm
  sh}>10^4$.}
\label{fig:tbar}
\end{figure}

We can now compute the temperature evolution of the IGM, including
shocks at both turnaround and virialization.  The results are shown in
Figure \ref{fig:tbar}.  We assume that the universe is reionized at
$z_r=8$; at this point the minimum mass for collapse increases
substantially.  The top panel shows the fraction of gas that is part
of the shocked phase, $F_{\rm sh}$, at a given redshift (solid curve)
as well as the fraction that is part of collapsed objects (dashed
curve).  The upper and lower dotted curves show $F_{\rm sh}$ before
reionization if we only include shocks with $T_{\rm sh}>10^4 \kel$ and
$10^5 \kel$, respectively.  The bottom panel shows the mass-averaged
temperature of the IGM.  The solid line is for shocked gas while
the dashed line is for collapsed gas (the temperature of each halo
is set to $T_{\rm vir}$, and we have used the Press-Schechter mass
function for the number density of halos at each mass).  The dotted
line shows the mean temperature from shocks if we include only those
with $T_{\rm sh}>10^4 \kel$ (i.e., ionizing shocks).  We set the
temperature of the rest of the gas to zero in each case.  The thermal
energy of the IGM is dominated by shocked gas at $z \ga 5$.  At the
present time, however, virialized objects determine the mean
temperature: so long as a fair fraction of the mass is in such
objects, their strong bounding shocks dominate over the IGM shocks
[compare to equation (\ref{eq:tcomp})].

\begin{figure}[t]
\plotone{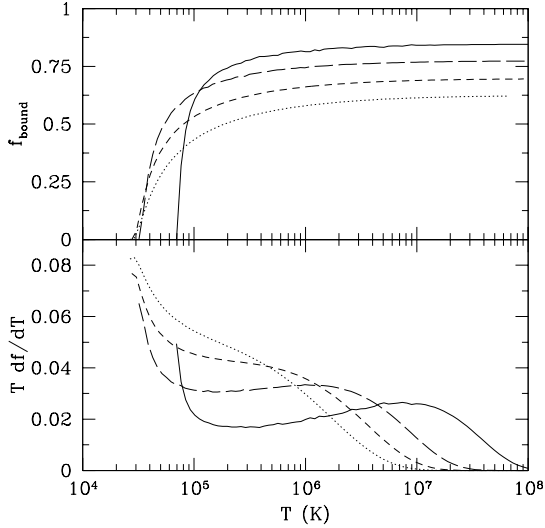}
\caption{The distribution of gas temperature in the shocked phase
  (\emph{bottom panel}) and the fraction of nominally shocked gas that
  is bound to collapsed halos (\emph{top panel}).  Solid, long-dashed,
  short-dashed, and dotted curves are for $z=0,\,1,\,2$, and $3$,
  respectively.  All assume $\delta_x=1.06$ and $\eta=5/16$.}
\label{fig:zlow}
\end{figure}

We see that the fraction in shocks is much larger than that in collapsed
objects at $z \ga 3$, except in a short window near to $z_r$.  Perhaps
surprisingly, the fraction of intergalactic shocked gas actually decreases
at low redshifts in our model as more and more mass is incorporated into
collapsed objects.  The bottom panel of Figure \ref{fig:zlow} shows the
temperature distribution function at several redshifts.  The solid,
long-dashed, short-dashed, and dotted lines show results for $z=0$, 1, 2,
and 3, respectively.  The top panel shows the fraction of gas nominally at
each shock temperature that is actually part of bound subhalos.  The bound
fraction rises rapidly above the minimum mass and then approaches a
constant value at a given redshift.  The asymptotic value is less than
unity because of the Jeans mass: a non-negligible fraction of the accreted
dark matter is bound up in potential wells too small to retain gas.

It is clear that the characteristic temperature of the shocked gas
declines rapidly with redshift; this is simply because the
characteristic mass scale also declines rapidly.  At $z \sim 2$--$3$,
the nonlinear mass scale is still sufficiently close to the minimum
collapse mass that a significant fraction of the gas is contained in
shocks rather than collapsed objects.  By $z=0$, most of the shocked
gas has already entered collapsed objects.  An important consequence
of this is that, at $z \ga 2$, most of the shocks have $T \sim 10^5
\kel$.  Because this is close to $T_{\rm IGM}$, the typical Mach
number of shocks is only $\sim 3$.  We expect our prescription for
shocks occurring at turnaround to be too optimistic in this case,
because large asymmetries would be required to produce shocks in such
a low Mach number flow.  Thus, our model possibly overestimates the
importance of shocks between reionization and $z \sim 2$ (compare to
Figure 5 of \citealt{dave01-whim}).  Figure \ref{fig:zlow}
demonstrates that (with fixed $\delta_x$ and $\eta$) our model does
\emph{not} accurately characterize the importance of shocks with Mach
numbers smaller than 10.  The fit may be improved by increasing the
density threshold at low Mach numbers or by decreasing $\eta$ (and
hence the shock velocity).  We are primarily interested in the strong
shocks prior to reionization where this problem does not arise, and so
we keep the approach simple and fix these two parameters to their
fiducial values.  It should also be noted that the finite resolution
of the simulations may also cause them to miss shocks (or smooth
them out through numerical viscosity) at these redshifts; \citet{ryu03} find
that shocks remain significant out to at least $z \sim 2$, in contrast
to the results of Dave et al (2001).  A more careful treatment of
shocks in the simulations may provide better agreement with our model.

We wish to apply our model to the high-redshift IGM in order to
consider the importance of shocks during that epoch.  The mean
temperature of the diffuse IGM is quite small before reionization (at
least if we neglect X-ray heating).  After recombination, the IGM
temperature remains coupled to the CMB temperature through Thomson
scattering off residual free electrons.  This mechanism becomes
inefficient at a redshift $z_{\rm dc}$ given by \citep{peebles93}
\bq 
1 + z_{\rm dc} \approx 132 (\Omega_b h^2/0.02)^{2/5},
\label{eq:zdc}
\eq
after which the temperature declines adiabatically as the universe
expands.  This means that $T_{\rm IGM} \ll T_{\rm CMB}$ at $z\sim20$,
at least until the first generations of sources produce sufficient
X-rays to heat the IGM (see \S \ref{cool} below and \citealt{chen03}).
The characteristic shock temperature, on the other hand, is much
larger than this at $z \la 20$.  Thus infall velocities are large and
we expect our model to provide a reasonable description.  Figure
\ref{fig:tbar} shows that, before reionization, the thermal energy of
the IGM is dominated by large-scale structure shocks rather than
virialized objects.  The disparity gets larger at high redshifts.
Essentially, the threshold for halo collapse, $M_{\rm fil}$, is near
or even well above the nonlinear mass scale at these epochs.  The
amount of mass in \emph{collapsing} gas is large compared to that in
collapsed halos, so it dominates the temperature of the IGM.

\begin{figure}[t]
\plotone{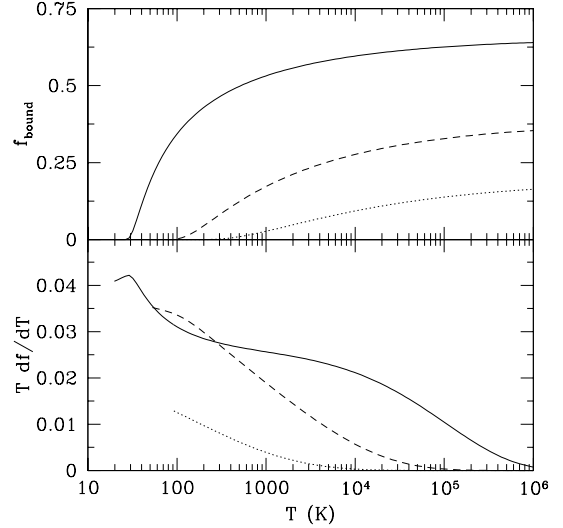}
\caption{The distribution of gas temperatures in the shocked phase
  (\emph{bottom panel}) and the fraction of nominally shocked gas that
  is bound to collapsed halos (\emph{top panel}).  Solid, dashed, and
  dotted curves are for $z=10,\,20$, and $30$, respectively.  All
  assume $\delta_x=1.06$ and $\eta=5/16$.}
\label{fig:zhigh}
\end{figure}

Figure \ref{fig:zhigh} shows the temperature distribution for $z=10$,
20, and 30.  Again, the characteristic temperature increases rapidly
with cosmic time.  At the highest redshifts, $df/dT$ decreases
smoothly with temperature and there is no characteristic temperature
peak.  This occurs because the nonlinear mass scale is below the
filter mass at these epochs.  This is also reflected in the bound
fraction.  This is small at high redshifts simply because the fraction
of gas bound to \emph{any} halo is small.

Note that these figures do not include radiative cooling. While
cooling is quite inefficient in the local IGM, it does become
important at higher redshifts because the characteristic density
increases and the cooling time decreases.  Before reionization,
cooling of regions with $T \ga 10^4 \kel$ is quite rapid; however,
cooling in weaker shocks is negligible because hydrogen remains
neutral and the gas is metal-free.  Moreover, the infalling gas is
continually accelerated by gravity, and our picture of a single shock
at turnaround is most likely too simple.  In reality, a complex
distribution of shocks around and inside filaments and sheets may
repeatedly shock the infalling gas, ``refreshing'' the energy that has
been lost to radiative cooling (c.f. \citealt{ryu03} at lower
redshifts).  We therefore expect the curves in the bottom panel of
Figure \ref{fig:tbar} to decrease somewhat if radiative cooling is
included.  In any case, most of the applications that we consider are
concerned with the total energy processed through the shocked gas,
which does not depend on subsequent cooling mechanisms.

\vspace{1cm}

\section{Effects of High-Redshift Shocks}
\label{dyneffects}

\subsection{Shock-Accelerated Particles at High-Redshifts}
\label{ptleacc}

We first consider cosmic ray acceleration in high-redshift shocks.
Particle acceleration is known to be efficient in high-temperature
collisionless shocks \citep{bell78,bland-ost}.  While rare at high
redshifts, such shocks still contain a substantial amount of energy
(see Figure \ref{fig:tbar}).  Even if only a small fraction of this
energy is transferred to cosmic rays, their energy density can still
exceed the thermal energy density of the unshocked, neutral IGM.  We
first show that ionizing shocks are collisionless, following
\citet{waxman01}.  We will only consider such shocks in this section
because acceleration is unlikely to be efficient if most of the
thermal energy is carried by neutral particles. For ionized shocks, we
wish to compare the ion-ion collision frequency,
\bq 
\nu_{ii} \approx 3 \times 10^{-9} \, \delta \, \left( \frac{T_{\rm
    sh}}{10^4 \kel} \right) \ \left( \frac{\Omega_b h^2}{0.02} \right)
\left( \frac{1+z}{20} \right)^3 \secinv,
\label{eq:nuii}
\eq
to the growth time for plasma instabilities, $\nu_g = C (v_s/c)
\nu_{pi}$, where the plasma frequency is
\bq
\nu_{pi} \approx 10 \, \delta^{1/2} \left( \frac{\Omega_b
  h^2}{0.02} \right)^{1/2} \left( \frac{1+z}{20} \right)^{3/2}
\secinv. 
\label{eq:nupi}
\eq
Here $C$ is a factor of order unity that depends on the detailed
plasma physics.  The shock is collisionless if $\nu_g > \nu_{ii}$, or
if
\bq 
C > 3 \times 10^{-6} \, \delta^{1/2} \, \left( \frac{T_{\rm
    sh}}{10^4 \kel} \right)^{-2} \left( \frac{\Omega_b h^2}{0.02}
\right)^{1/2} \left( \frac{1+z}{20} \right)^{3/2}.
\label{eq:collcond}
\eq
We therefore expect particle acceleration to be efficient
\citep{bland-eichler}, though we note that the conditions are
sufficiently different from the best studied cases (supernova
blastwaves and cluster shocks) that the properties of the acceleration
mechanism are not well-constrained observationally.

We assume that plasma instabilities induce a magnetic field containing
a fraction $\xi_B$ of the thermal energy in equipartition.  In such a
situation, particles are accelerated to high energies by scattering
off fluctuations in the magnetic field \citep{bell78,bland-ost}.  The
acceleration time is $t_{\rm acc} \sim r_L c/v_s^2$, where
$r_L$ is the Larmor radius of the particle.  The maximum
Lorentz factor $\gamma_{\rm max}$ is fixed by the condition 
$t_{\rm acc} \sim t_{\rm IC}$, where $t_{\rm IC}$ is the inverse
Compton (IC) cooling time of the particle (the dominant cooling
mechanism at high redshifts).  The requirement is \citep{loeb00,keshet03}
\begin{eqnarray}
\gamma_{\rm max} & \sim 10^3 &
\left( \frac{\xi_B }{0.01} \right)^{1/4} \left( \frac{\delta}{10}
\right)^{1/4} \left( \frac{\Omega_b h^2}{0.02} \right)^{1/4} 
\nonumber \\ 
& & \times \left(
\frac{1+z}{20} \right)^{-5/4} \, \left( \frac{T_{\rm sh}}{10^4 \kel}
\right).
\label{eq:gammamax}
\end{eqnarray}

In fact, even those electrons well below this threshold will lose all
their energy to IC emission within a Hubble time so long
as 
\bq
\gamma + 1 > 0.43 \left( \frac{10}{1+z} \right)^{5/2}.
\label{eq:gammaic}
\eq
We can therefore assume that all of the energy entering the
accelerated electron component is transformed into IC radiation (which
is a much more efficient cooling channel than synchrotron losses at
high redshifts given that $\xi_B\le 1$).  Assuming a strong shock, the
resulting spectrum is flat ($J_\nu \propto \nu^{-1}$;
\citealt{rybicki}).  We estimate the energy density as in
\citet{loeb00}:
\bq
\nu u_\nu \sim \frac{1}{2 \ln(2 \gamma_{\rm max})} \times \xi_e \times
    {3 \over 2} n_b k \bar{T}_{\rm sh}, 
\label{eq:ubkgd}
\eq
where $\bar{T}_{\rm sh}$ mass-averaged temperature of the IGM and
$\xi_e$ is the fraction of thermal energy behind the shock that is
used to accelerate the electrons.  The factor $1/2$ enters because
each decade in electron Lorentz factor corresponds to two decades in
photon frequency.  Note that $\bar{T}_{\rm sh}$ should include only
those shocks that accelerate electrons (i.e., those with $T_{\rm
sh}>10^4 \kel$).  We show this quantity as the dotted line in the
bottom panel of Figure \ref{fig:tbar}.  Note that, by $z \sim 20$,
most of the thermal energy is contained in such shocks.  At the
emission redshift, the IC spectrum will extend between
$(\varepsilon_{\rm CMB},\gamma_{\rm max}^2 \varepsilon_{\rm CMB})
\approx (10^{-3},10^3) \times (1+z) \eV$.  The background flux is then
$J_\nu = c/4\pi \times 
u_\nu$:
\begin{eqnarray}
J_{\nu,21}^{\rm IC} & \sim \frac{10^{-4}}{\ln(2 \gamma_{\rm
    max}/1000)} & \, \left(
\frac{\xi_e}{0.05} \right) \, \left( \frac{\Omega_b h^2}{0.023}
\right) \left( \frac{1+z}{20} \right)^3 \nonumber \\
& & \times \left( \frac{\bar{T}_{\rm
    sh}}{100 \kel} \right) \, \left( \frac{\nu_{\alpha}}{\nu}
\right), 
\label{eq:jbkgd}
\end{eqnarray}
where $J_{\nu,21}=J_\nu/(10^{-21} \junits)$ and $\nu_{\alpha}$ is
the frequency of the hydrogen Ly$\alpha$ transition. 

We thus find that particle acceleration at large-scale structure
shocks creates a diffuse radiation background with a flat spectrum
ranging from radio to soft X-ray energies.  Interestingly, such
particles may be the first source of high energy ($\ga 1$ eV)
radiation in the universe after cosmological recombination.  We expect
particle acceleration to begin once the first structures with $T_{\rm
sh} \ga 10^4 \kel$ turn around.  This occurs well before $z \sim 30$
(albeit only rarely).  In contrast, simulations show that the first
stars form at $z \sim 20$--$30$ \citep{abel,bromm02}; this is both
because the star-forming regions must have fully collapsed and because
subsequent cooling is regulated by H$_2$ formation and is quite slow.
We consider some of the consequences of this radiation background in
the next section.

In addition to electrons, shocks will deposit a fraction $\xi_p \ga
\xi_e$ of their energy in cosmic ray protons.  Such protons do not
cool through IC radiation; in fact, they become an important pressure
component of the shock.  This component may actually have a higher
energy density than the diffuse IGM (averaged over the entire
universe):
\bq 
\frac{u_{\rm CR}}{u_{\rm IGM}} \sim \left(
\frac{\xi_p}{0.1} \right) \ \left( \frac{\bar{T}_{\rm sh}}{100 \kel}
\right) \left( \frac{20}{1+z} \right)^2.
\label{eq:crden}
\eq
Note that the shocks continue to accelerate protons until the proton
Larmor radii exceed the shock thickness.  We therefore expect the high
energy tail of the distribution (which contains a substantial fraction
of the total energy for a strong shock, see equation [\ref{eq:ubkgd}])
to escape into the diffuse IGM and fill out the voids uniformly.
However, the short diffusion time of the relativistic ions in the
unmagnetized IGM implies that their pressure has a weak effect on the
assembly of gas during galaxy formation.

\subsection{Diffuse Radiation}
\label{cool}

One way for the thermal shock energy to be transformed into photons is
through IC emission by accelerated electrons, as described in the
previous section.  Another is direct radiative cooling.  Figure
\ref{fig:tbar} shows that the thermal energy budget of the IGM is
dominated by shocked regions before reionization.  We therefore expect
the background radiation field from cooling gas to be dominated by
shocked regions as well (provided that the cooling time is short in
these regions, which is true so long as $T_{\rm sh} \ga 10^4 \kel$).
Cooling occurs principally through IC emission, line transitions, and
free-free emission.  For the moderately overdense regions with $T \sim
10^4$--$10^5 \kel$ in which we are most interested, line cooling
typically dominates.  IC emission from thermal electrons produces
extremely low-energy photons, so we will not consider it further
here. 

Most of the cooling occurs through hydrogen and helium line emission,
which the background spectrum at $E<13.6 \eV$ .  The flux at a given
frequency $\nu$ is the sum of fluxes from those lines that redshift to
$\nu$ at the appropriate cosmic time, after adjusting for radiative
transfer in the IGM \citep{haiman97}.  We can roughly estimate the
background flux by
\begin{eqnarray}
J_{\nu,21}^{\rm line} & \sim & \frac{c}{4\pi} \times
f_{\rm line} \frac{3}{2} k n_b \bar{T}_{\rm sh} \times \frac{2}{3
  (1+z) \nu}
\label{eq:linebkgd} \\
\, & \sim & 2 \times 10^{-3} f_{\rm line} \, \left( \frac{\bar{T}_{\rm
      sh}}{100 \kel} \right) \ \left( \frac{\nu_\alpha}{\nu} \right)
      \ \left( \frac{1+z}{20} \right)^2,
\nonumber
\end{eqnarray}
where $f_{\rm line}$ is the fraction of the thermal energy lost to
this line and again $\bar{T}_{\rm sh}$ should only include ionizing
shocks.  Note that $\bar{T}_{\rm sh}$ should actually be evaluated at
the $z'>z$ appropriate to the specified frequency.  For example, to
evaluate the contribution to $J_\nu(\nu_\alpha)$ at a redshift $z$
from Ly$\beta$ emission, we would set $1+z'=(\nu_\beta)/(\nu_\alpha)
\times (1+z)$.

Next we can estimate the spectrum of free-free radiation as follows.
First we approximate the bremsstrahlung emissivity $n_e^2
\varepsilon^{ff}_\nu$ as flat below a cutoff freqency given by $h
\nu_c \sim k T$.  Thus the energy density at a particular frequency is
simply proportional to the amount of gas that has been shock-heated
above the corresponding temperature.  Each parcel of shocked gas
radiates over its cooling time (where we must include all emission
mechanisms).  We find
\begin{eqnarray}
J_{\nu,21}^{ff} & \sim & \frac{c}{4\pi} \times \frac{3}{2} n_b k T
f_{\rm sh}(>T) \times \frac{\varepsilon^{ff}_\nu}{\varepsilon(T)}
\label{eq:jnuff} \\
& \sim & f_{\rm sh}(>T) \ \left( \frac{T}{10^5 \kel} \right)^{1/2} \ \left(
\frac{10^{-23} \coolfcn}{\varepsilon(T)} \right) \nonumber \\
& & \times \left(
\frac{\Omega_b h^2}{0.023} \right) \ \left( \frac{1+z}{20} \right)^3,
\nonumber
\end{eqnarray}
where $\varepsilon(T)$ is the cooling function (including all
processes) and $f_{\rm sh}(>T)$ is the fraction of gas that has been
shock-heated to $k T > h \nu$.  In the first line, the second factor
is the thermal energy in the shocks warm enough to to emit at the
relevant frequency, while the third factor is the fraction of energy
lost to free-free emission at the appropriate frequency.  (We have
assumed here that the cooling time is smaller than the Hubble time; if
not, the flux decreases by the appropriate factor.)  The universe is
essentially transparent to photons with $\nu < \nu_\alpha$
(corresponding to a temperature threshold $T_{\rm sh} \ga 10^5 \kel$),
while above this threshold the radiation interacts with the IGM gas.
We show $f_{\rm sh}(>10^5 \kel)$ as the lower dotted curve in Figure
\ref{fig:tbar}: such shocks are extremely rare until $z \sim 10$.  We
see that IC emission from accelerated particles will dominate the
high-energy background radiation field at X-ray energies, but the
free-free emission may be significant at UV energies for $z \la 15$.

While these estimates are quite rough, it is clear that (neglecting
emission from protogalaxies) the radiation spectrum from a few eV to
the ionization threshold of hydrogen will be dominated by free-free
line emission from cooling objects, with a higher energy tail due to
IC emission from accelerated electrons.  This diffuse background may
have dynamical effects on structure formation.  We list several of
these below.

\begin{itemize}
\item \emph{Ionizing background:}  We can estimate the ionization
  fraction due to the IC background as $x_e \sim f_i u_{\rm
  IC}/(n_b E_i)$, where $f_i$ is the fraction of the radiation
  background that goes into ionizations, $E_i$ is the mean energy lost
  per ionization, and $u_{\rm IC} = \xi_e \times (3/2) n_b k
  \bar{T}_{\rm sh}$ is the total energy density in the radiation
  field.  Note that $f_i \la 1/2$ accounts for both those photons below the
  ionization threshold and the fraction of energy in high-energy
  photons that goes into heating rather than ionization.  The
  resulting electron fraction is
\begin{eqnarray}
x_e & \sim 3 \times 10^{-4} & \ \left( \frac{f_i}{0.5} \right) \, \left(
  \frac{\xi_e}{0.05} \right) \nonumber \\ 
& & \times  \left( \frac{13.6 \eV}{E_i} \right) \, \left( \frac{\bar{T}_{\rm
  sh}}{10^3 \kel} \right).
\label{eq:ionbkgd}
\end{eqnarray}
The residual electron fraction from cosmological recombination is $x_e
\sim 2 \times 10^{-4}$ with our assumed cosmological parameters.  We
thus find that ionization from IC emission could only become important
at $z \la 15$ and even then only if star formation is quite
inefficient and hence subdominant.  On the other hand, a significant
fraction of the IC photons have keV energies, and these are able to
propagate much farther than lower-energy stellar ionizing photons.

\item \emph{X-ray heating:} An additional effect of the high energy
  photons is to heat the diffuse IGM gas.  We can estimate the
  resulting temperature by $T_X \sim f_h \xi_e
  \bar{T}_{\rm sh}$, where $f_h$ is the fraction of IC energy that
  goes into heating the gas.  Comparing to the temperature of the gas
  $T_{\rm IGM}$ (without additional heating, see \S \ref{tempevol}
  above), we find 
\begin{eqnarray}
\frac{T_X}{T_{\rm IGM}} & \sim 10 & \ \left( \frac{f_h}{0.5} \right) \,
  \left( \frac{\xi_e}{0.05} \right) \nonumber \\
& & \times \left( \frac{\bar{T}_{\rm
  sh}}{10^3 \kel} \right) \, \left( \frac{1+z}{10} \right)^{-2}.
\label{eq:txray}
\end{eqnarray}
Note that $T_X \sim T_{\rm CMB}$ at $z=10$ with the parameter choices
shown.  Even though only a small fraction of the shock energy goes
into particle acceleration, the diffuse IGM gas is so cold that the
heat input can be significant, even at $z \sim 20$.  The temperature
of the IGM ultimately determines the minimum mass of halos that can
form and is potentially measurable with the 21 cm transition of
neutral hydrogen (see \S \ref{21cm} below).

\item \emph{Effects on H$_2$ formation:} Without metals, cooling to
  the low temperatures relevant for star formation is controlled by
  H$_2$ \citep{abel,bromm02,barkana01}.  Radiative feedback can be
  significant in the formation of molecular hydrogen; ultraviolet
  photons in the Lyman-Werner band ($11.26$--$13.6 \eV$) can
  dissociate H$_2$ \citep{haiman97,machacek01}, while X-rays may
  catalyze its formation by increasing the density of free electrons
  \citep{haiman00}, although this effect remains controversial
  \citep{machacek03,oh03-entropy}.  Photodissociation of H$_2$ becomes
  important when $J_{21} \ga 10^{-3}$ in the UV band; we expect that
  this could become significant at $z \sim 20$ in our model [see
  equation (\ref{eq:linebkgd}).  The IC component has $J_{21} \sim
  10^{-6} ({\rm keV}/E)$, sufficiently far below the UV flux that
  photodissociation should dominate \citep{haiman00}.

\item \emph{Wouthuysen-Field Effect:} There has been much interest
  recently in observing neutral gas at high-redshifts through the 21
  cm spin-flip transition of \ion{H}{1} (see \S \ref{21cm} below).
  For this to be possible, the spin temperature $T_S$ of the hyperfine
  transition must decouple from $T_{\rm CMB}$.  In the diffuse IGM,
  decoupling occurs through the Wouthuysen-Field effect
  \citep{wout,field58}. In this process a hydrogen atom absorbs and
  re-emits a Ly$\alpha$ photon; the selection rules allow mixing of
  the two hyperfine states during this process.  The Wouthuysen-Field
  effect becomes significant when $J_{\nu,21}(\nu_\alpha) \ga 1$
  \citep{mmr}.  This requires $\bar{T}_{\rm sh} \ga 10^4 \kel$ in
  equation (\ref{eq:linebkgd}), which does not occur until $z \la 8$.
  We therefore find that, in our model, radiation from shocks cannot
  decouple $T_S$ and $T_{\rm CMB}$.  Stellar radiation is
  \emph{required} in order to observe the diffuse gas through the 21
  cm transition (see \citealt{ciardi03,chen03}).

\end{itemize}

Of course, both IC and cooling radiation come from discrete sources
and are not truly diffuse backgrounds.  This is especially true at
high redshifts, where shocks with $T_{\rm sh} > 10^4 \kel$ are quite
rare.  We therefore expect the above effects to be significantly
larger near to the shocks themselves.  Ionization and feedback effects
on H$_2$ would be considerably larger in these regions, which is
especially important because both shocks and collapsing structures
preferentially form in the most highly biased regions.  In principle,
the cooling radiation from individual shocks could be imaged.
Unfortunately, while the luminosities are large, the sources are quite
extended and the surface brightness is extremely small.  Even the
\emph{James Webb Space Telescope} would require hundreds of hours of
integration to see an individual system at $z>10$.

\subsection{Gas Stripping from Minihalos}
\label{strip}

At low and moderate redshifts, large-scale structure shocks have
complex morphologies \citep{ryu03} that surround and interpenetrate
sheets and filaments.  Nearly all of the gas that accretes smoothly
onto the structures is shocked; however, it is less clear whether gas
that has cooled onto galaxies or even galaxy groups is affected by the
shocks.  The galactic interstellar medium is much denser than the
surrounding IGM, which dramatically decreases the strength of the
accretion shocks felt by the interstellar medium.  Moreover, these
objects have substantial magnetic fields that may deflect gas in the
external medium around the object, rather than allowing it to collide
and shock the interstellar medium of the galaxy.

However, these mechanisms do not apply to gas in ``minihalos'' at high
redshifts.  Such objects have virial temperatures $T_{\rm vir}<10^4
\kel$, placing them below the threshold to excite atomic hydrogen
transitions.  In the absence of molecular hydrogen, they are therefore
unable to cool efficiently and cannot collapse after virialization,
remaining as moderately overdense ($\delta \sim 200$) gaseous halos in
the IGM (see \citealt{barkana01} and references therein).  In terms of
their internal structure, such halos are thus more similar to local
galaxy clusters than to galaxies, except that they likely have
negligible magnetic fields and no gas substructure.

Thus, these halos have little protection from the ram pressure of the
surrounding medium.  The minihalo gas will then be stripped to a
radius $R$ determined by the condition $\rho_{\rm sh} v_{\rm sh}^2 =
\rho_{\rm mh}(R) \sigma_{\rm mh}^2$, where $\rho_{\rm sh}=\delta_{\rm
sh}\bar{\rho}$ is the mean density in the shocked region, $\rho_{\rm
mh}(R)=\delta_{\rm mh}(R)\bar{\rho}$ is the density at a given point
within the minihalo, and $\sigma_{\rm mh}$ is the velocity dispersion
of the minihalo.  The condition is approximately
\bq
\delta_{\rm mh}(R) \approx \frac{\gamma}{2 \eta} \left( \frac{T_{\rm
    sh}}{T_{\rm vir}} \right) \delta_{\rm sh}.
\label{eq:ram}
\eq 
Given a density profile for the minihalo gas, this prescription allows
us to compute the fraction of minihalo gas that is lost as it travels
through large-scale structure shocks.  We consider two possible
density profiles: a singular isothermal sphere (SIS) and isothermal
gas inside of a dark matter halo with the structure found by
\citet[][hereafter NFW]{nfw}.  We parameterize our results in terms of
$f_{\rm ret}$, the fraction of the initial minihalo gas mass retained
by the minihalo after ram pressure is taken into account.  For an SIS,
the criterion is particularly simple: $f_{\rm ret} = R/r_{\rm vir}$.
For isothermal gas in an NFW halo,
\bq
f_{\rm ret}(R) = \frac{\int_0^{R/r_{\rm vir}} du \, u^2 (1+c
u)^{g(c)/u}}{\int_0^1 du \, u^2 (1+cu)^{g(c)/u}},
\label{eq:nfwiso}
\eq
where
\bq
g(c) = \frac{2}{\ln(1+c)-c/(1+c)}
\label{eq:gc}
\eq
and $c$ is the concentration parameter of the minihalo.  We choose $c$
from the fit to simulations given by \citet{bullock01}.

\begin{figure}[t]
\plotone{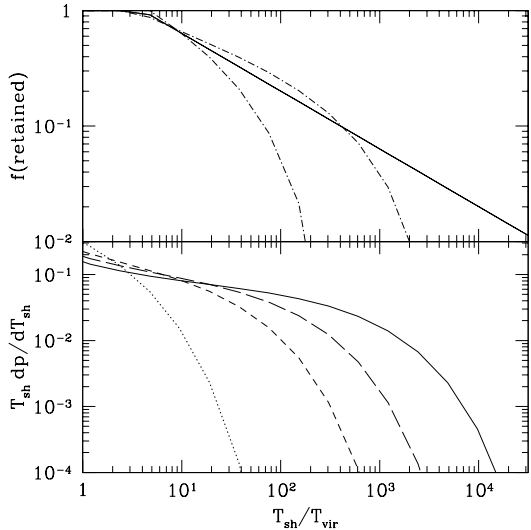}
\caption{\emph{Bottom panel:} The probability that a halo at $z=10$ is
contained within a shock of a given strength.  The solid, long-dashed,
short-dashed, and dotted lines are for halos with $T_{\rm
vir}=170,\,1080,\, 5030,\,$ and $10^5 \kel$.  \emph{Top panel:} The
fraction of mass retained in a halo after ram-pressure stripping.  The
solid line assumes the gas has an SIS distribution and the dot-dashed
lines assume that the gas is isothermal in an NFW halo.  In the latter
case, the left and right curves assume concentration parameters
appropriate for halos with $T_{\rm vir}=170$ and $10^5 \kel$,
respectively.  }
\label{fig:disruption}
\end{figure}

Figure \ref{fig:disruption} shows the effects on individual halos.
The bottom panel shows the probability that a halo at $z=10$ is
contained within a shock as a function of $T_{\rm sh}/T_{\rm vir}$.
The solid, long-dashed, short-dashed, and dotted curves assume $T_{\rm
vir}=170,\, 1080,\, 5030$, and $10^5 \kel$, respectively.  These
correspond to $M_h=6 \times 10^{4},\, 10^6,\, 10^7$, and $10^9 \msun$.
The smallest of the these masses is the filter mass at $z=10$ while
the largest is above the cooling threshold and is included only for
illustrative purposes.  We see that the probability for a halo to sit
inside a strong shock is extremely sensitive to its mass.  A
substantial fraction of halos below the nonlinear mass scale are
embedded in strong shocks, but very few halos above that scale are
associated with larger shocks because such structures are increasingly
rare.  The top panel shows the fraction of mass retained in a halo as
a function of $T_{\rm sh}/T_{\rm vir}$.  The solid line assumes the
gas has an SIS distribution ($f_{\rm ret}$ is then independent of halo
mass when expressed as a function of $T_{\rm sh}/T_{\rm vir}$) and the
dot-dashed lines assume that the gas is isothermal in an NFW halo.
The concentration parameters in this plot are for $T_{\rm vir}=170$
and $10^{5} \kel$ (left and right curves); this is the only aspect in
which the halo mass enters into this parameterization.  Stripping sets
in when the shock temperature is a few times the virial temperature of
the minihalo.  The SIS model is more concentrated and is therefore
more stable to mass loss.  Note that the NFW model has a finite
density core, so the entire minihalo can be stripped.  Comparing the
two panels, we see that a negligible fraction of mass would be lost
from halos near to the minihalo mass threshold, but a large fraction
could be lost from smaller halos.

\begin{figure}[t]
\plotone{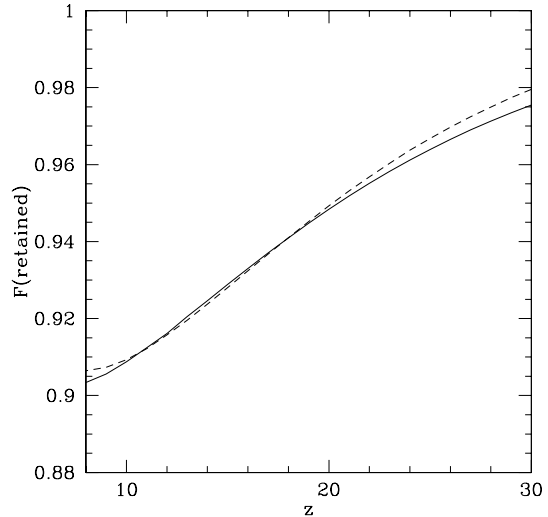}
\caption{Fraction of gas in collapsed objects that is retained after
  ram pressure stripping.  The solid curve assumes the minihalo
  profile is an SIS, while the dashed curve assumes isothermal gas in
  an NFW potential.}
\label{fig:fracstrip}
\end{figure}

Figure \ref{fig:fracstrip} shows the global effects of shock
disruption.  We show the fraction of mass in collapsed objects
(including both minihalos and halos that are able to cool) that
remains after ram pressure stripping is included.  The solid line
assumes that minihalos have SIS density profiles while the dashed line
assumes they are isothermal gas in an NFW potential.  The two density
profiles give very similar results because they differ substantially
only in the core (which rarely gets stripped).  We see that the global
effects of stripping are fairly small: only $\la 5\%$ of the gas is
stripped at $z \ga 20$ while approximately $10\%$ of the gas is
stripped at $z \sim 10$.  The importance of stripping increases with
cosmic time because, as the nonlinear mass scale increases, regions
with stronger and stronger shocks form and a larger fraction of
minihalos are contained in such regions.  Note that stripping is
concentrated in the most highly biased regions of the universe (where
strong shocks can form).  This is also where a large fraction of
ionizing sources reside; we would therefore expect minihalos to be
less important around such sources than previous estimates have
assumed (e.g., \citealt{barkana02,shapiro03}).  These estimates also
assume that minihalo gas is stripped only \emph{after} the minihalo
has formed.  In reality, many minihalos will enter large-scale
structure shocks during their own collapse.  Criterion (\ref{eq:ram})
suggests that stripping is more efficient if the initial density is
smaller.

Ram pressure is not the only way in which shocks can strip minihalos.
As they fall into these structures, the minihalo gas will pass through
the shock.  If the shock adds sufficient entropy to the gas, it can
escape the minihalo \citep{oh03-entropy}.  The stripping efficiency is
then determined by the ratio of the shock speed in the minihalo [which
is proportional to $\rho(R)^{-1/2}$] to the sound speed.  We find that
this mechanism is slighly less efficient than ram pressure stripping,
so the two mechanisms together will likely increase the amount of
stripped gas over that shown in Figure \ref{fig:fracstrip} by a small
factor.  Calibration of the precise level of these effects must await
numerical simulations that include a careful treatment of the shock
physics as well as broad enough dymanic range to resolve minihalos
while also encompassing a large, representative cosmological volume.

Finally, note that shocks may trigger star formation in minihalos.  If
the postshock temperature $T_{\rm sh} \ga 10^4 \kel$, the gas will be
able to cool through both line radiation and inverse-Compton
processes.  In this case, gas would remain bound to the minihalo and,
if cooling is sufficiently fast, some gas could even collapse to high
enough densities to form stars \citep{mackey03,cen03-shock}.  In
reality, the dense central regions of the minihalos may collapse to
form stars while the outer layers are stripped by the shock.
Disentangling these two effects requires hydrodynamic simulations of
the shocking process.  Triggered star formation is likely to be less
efficient in large-scale structure shocks than in supernova blastwaves
because the characteristic velocities are smaller.

\section{Observational Signatures of High-Redshift Shocks}
\label{observe}

\subsection{Optical Depth to the CMB}

\begin{figure}[t]
\plotone{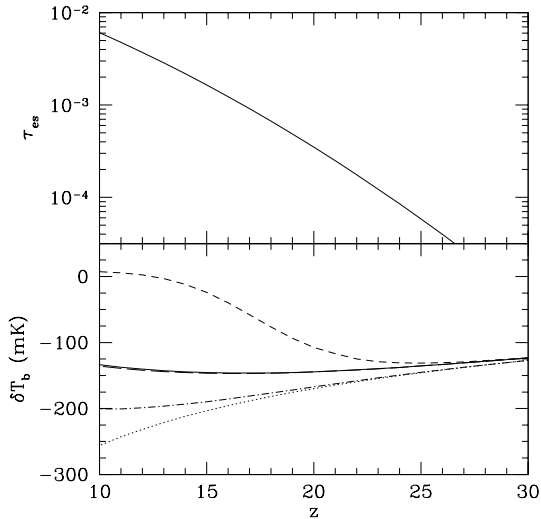}
\caption{ \emph{Top panel:} The optical depth to electron scattering
$\tau_{\rm es}$ in our model, assuming that all gas associated with
bound objects and/or turnaround shocks of $T>10^4 \kel$ remains
ionized.  \emph{Bottom panel:} The mean 21 cm brightness temperature
decrement $\delta T_b$ of the IGM relative to the CMB.  The dotted
curve neglects all heat sources (including gravity).  The dot-dashed
curve includes virial shocks.  The solid and long-dashed curves
include heating at turnaround shocks; the latter assumes that all gas
in shocks with $T>10^4 \kel$ remains ionized and is almost hidden by
the solid curve.  The short-dashed curve also includes X-ray heating
through the IC background from shock-accelerated electrons.}
\label{fig:global}
\end{figure}

One way to observe the effects of gravitational heating is through its
effects on the CMB.  Large-scale shocks with $T \ga 10^4 \kel$ will
ionize gas that passes through them.  CMB photons will scatter off the
resulting free electrons, smoothing out the anisotropies.  The
cumulative optical depth to electron scattering $\tau_{\rm es}$
between redshifts $z_l$ and $z_h$ is
\bq
\tau_{\rm es} = \int_{z_l}^{z_h} dz \sigma_T n_e(z) c \frac{dt}{dz},
\label{eq:taues}
\eq
where $\sigma_T$ is the Thomson scattering cross section.  We compute
the optical depth by assuming that all gas associated with shocks of
$T>10^4 \kel$ at a given redshift is ionized.  The result is shown in
the top panel of Figure \ref{fig:global}; $\tau_{\rm es} < 1\%$ and is
unlikely to contribute substantially to the measured optical depth,
particularly if reionization occurs at high redshift.  By neglecting
recombination of the gas, we actually provide an upper limit to the
optical depth due to IGM shocks.  Recombination will be significant
within the overdense shocked regions, unless the gas is shocked many
times as it falls into the potential well of an object.

\subsection{21 cm Radiation}
\label{21cm}

Another way to identify shocks is through their effect on the
global temperature distribution of the IGM, which can be measured
with the 21 cm transition.  The brightness temperature increment of
the 21 cm transition from neutral gas relative to the CMB is
\citep{field58,mmr} 
\begin{eqnarray}
\delta T_b & = 26 \mkel & \, \left( \frac{\Omega_b h^2}{0.023}
\right) \left( \frac{0.3}{\Omega_m} \, \frac{1+z}{10} \right)^{1/2}
\nonumber \\
& & \times  \int dT_S \frac{df}{dT_S} \, \frac{T_S-T_{\rm CMB}}{T_S},
\label{eq:dtb}
\end{eqnarray}
where $T_S$ is the spin temperature of the hydrogen and the integral
extends over all \emph{neutral} gas.  Note that if $T_S \gg T_{\rm
CMB}$, the signal saturates in emission, but if $T_S \ll T_{\rm CMB}$
the magnitude of the absorption increases like $T_S^{-1}$.

The spin temperature remains locked to the CMB temperature unless (1)
the background Ly$\alpha$ photon field is large enough for the
Wouthuysen-Field mechanism to become efficient (see \S \ref{cool}
above) or (2) the number density of hydrogen is large enough for
collisions to decouple the two \citep{field58}.  If these processes
operate,
\bq
T_S = \frac{T_{\rm CMB} + y_\alpha T_\alpha + y_c T_K}{1 + y_\alpha +
  y_c},
\label{eq:tspin}
\eq 
where $T_\alpha$ is the Ly$\alpha$ color temperature ($T_\alpha=T_K$
under a wide variety of conditions; \citealt{field59b}) and $T_K$ is
the kinetic temperature of the gas.  The coupling constant
$y_\alpha=P_{10} T_\star/(A_{10} T_\alpha)$, where $P_{10} =1.3 \times
10^{-12} J_{\nu,21}(\nu_\alpha)$ is the indirect de-excitation rate of
the triplet level due to the absorption of a Ly$\alpha$ photon
followed by decay to the singlet level, $T_\star=0.068 \kel$ is the
excitation temperature of the hyperfine level, and $A_{10}=2.85 \times
10^{-15} \secinv$ is the spontaneous emission coefficient of the
transition.  The coupling from collisions is $y_c=C_{10}
T_\star/(A_{10} T_\alpha)$, where $C_{10}$ is the collisional
de-excitation rate of the triplet hyperfine level and has been
calculated for $T_K<10^3 \kel$ by \citet{allison} (we assume $C_{10}
\propto T_K^{0.33}$ for higher temperatures).  

We first assume that $T_S=T_K$ throughout the universe (i.e., there is
a strong Ly$\alpha$ background).  We showed in \S \ref{cool} that this
requires stellar (or quasar) radiation, because the background from
shocks themselves is insufficient.  In the absence of heat sources,
the IGM will appear as a smooth \emph{absorption} feature against the
CMB [$df/dT_S=\delta(T-T_{\rm IGM})$].  The signal is shown by the
dotted curve in the bottom panel of Figure \ref{fig:global}; we see
$|\delta T_b| \ga 100 \mkel$, decreasing rapidly as redshift decreases
because $T_K \propto (1+z)^2$.  This signal is sufficiently strong to
be easily observable by the next generation of low-frequency radio
telescopes.  However, structure formation will remove gas from the
cold IGM, decreasing the overall absorption signal.  First, we must
remove gas that is attached to bound objects.  In this case the
integral in equation (\ref{eq:dtb}) extends over the mass function of
collapsed objects $dn/dM$.  The result is shown in Figure
\ref{fig:global} as the dot-dashed line.  We see that $\delta T_b$
changes by only a small amount for $z \ga 15$.

Next, we include (in addition to collapsed objects) gas that has been
shock-heated by large-scale structure.  This is shown by the solid
line in Figure \ref{fig:global}, assuming that all gas remains
neutral.  (The long-dashed line shows the signal if gas associated
with $T>10^4 \kel$ shocks remains ionized and does not emit; it is
nearly completely obscured by the solid curve.)  Because $f_{\rm sh}
\gg f_{\rm coll}$ at high redshifts, the absorption signal flattens
out at much higher redshifts; even at $z=20$, $|\delta T_b|$ is
smaller by about $30 \mkel$.  In principle, shock-heating at high
redshifts could therefore be detected by its effects on the radio
spectrum.  However, other heat sources, such as X-rays from the IC
background generated by particles accelerated at the shocks (see \S
\ref{cool}) or from supernovae or quasars, will increase the
temperature of the low-density IGM and may cause significant
contamination with this signal \citep{chen03}.  For example, the
short-dashed line shows the signal assuming the heat input from
equation (\ref{eq:txray}), which includes IC emission from
shock-accelerated electrons.  Even this relatively gentle heating
rapidly decreases the amount of absorption, and it actually turns into
emission by $z \sim 13$.  We would therefore require an early stellar
Ly$\alpha$ background (unaccompanied by X-ray heating) for significant
absorption to be possible.  Note that the treatment of weak shocks has
a substantial effect on this estimate, so our model may not be
quantitatively accurate.  \citet{gnedin03} have recently simulated
this absorption era; they find that shock heating decreases $|\delta
T_b|$ by an even larger amount than we estimate. This indicates that
weak shocks preceding turnaround may heat the diffuse IGM by a small
amount before reionization and that we \emph{underestimate} their
effects (unlike at moderate redshifts).

The above discussion of the globally-averaged signal ignores the fact
that the temperature and density distributions are highly
inhomogeneous.  The absorption and emission signal therefore
fluctuates across the sky, and one can also search for these
fluctuations \citep{scott,mmr,zald03,loeb03}.  Because the emission
signal is independent of $T_S$ for $T_S \gg T_{\rm CMB}$, it
essentially depends only on the mass of gas in a given volume that
emits.  Let us now consider a sufficiently early epoch when the
Ly$\alpha$ background is small, so that $T_S=T_{\rm CMB}$ in the
diffuse IGM.  \citet{iliev} have shown that, because collisional
coupling is still effective in dense gas, minihalos provide a
fluctuating emission signal that may be observable on large scales
(see also \citealt{iliev03}).  Here we point out that, because the
shocked phase contains so much more mass than minihalos, the emission
signal is probably much larger than \citet{iliev} have estimated.

\begin{figure}[t]
\plotone{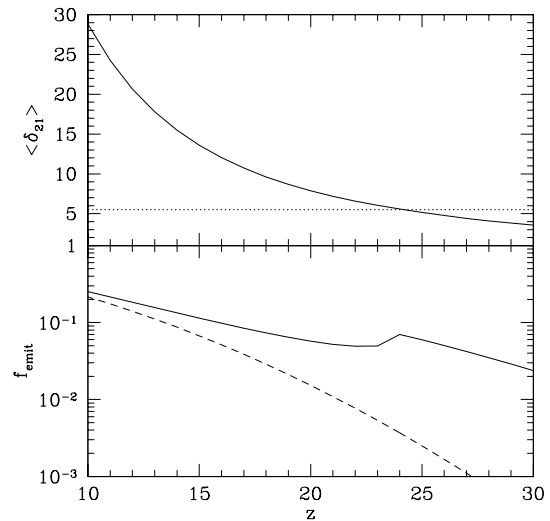}
\caption{\emph{Top panel:} The critical density $\langle\delta_{21}\rangle$
  for coupling between the spin and kinetic temperatures of
  \ion{H}{1}, averaged over the temperature of the shocked phase.  The
  dotted line shows $\delta=5.55$, the nominal density at turnaround.
  \emph{Bottom panel:} The fraction of gas able to emit.  The solid
  line is for the shocked phase (see text); the dashed line includes
  only the collapsed objects.}
\label{fig:21cmemit}
\end{figure}

In the case of minihalos, the density is so large that $y_c \gg 1$
throughout the halo.  Thus $T_S=T_K$ is a good approximation in these
systems.  In shocks, however, the densities are much
smaller and not all of the gas may have $T_S > T_{\rm CMB}$.  Given
$z$ and $T_K$, we first define $\delta_{21}$, the overdensity for
which $y_c=1$ and above which collisional coupling becomes efficient.
Note that (at fixed temperature) $\delta_{21}$ increases as $z$
decreases because the mean density also decreases.  We then let $\langle
\delta_{21} \rangle$ be the critical overdensity for collisional coupling
to be effective, averaged over the temperature distribution of the
shocked gas:
\bq
\langle \delta_{21}(z) \rangle = \int d T_K \ \frac{df}{dT_K}
\delta_{21}(z,T_K) 
\label{eq:d21avg}
\eq
The resulting values are shown by the solid line in the
top panel of Figure \ref{fig:21cmemit} for our shock model.  The
dotted line shows the overdensity at turnaround in the spherical
model; at $z \ga 25$, all shocked gas emits strongly.  Note that
$\langle \delta_{21} \rangle \ll 200$ for all redshifts shown.

We wish to find the fraction of collapsing gas that has $\delta >
\langle \delta_{21} \rangle$.  In the spherical collapse model, the
overdensity at turnaround is $\delta_{\rm ta}=5.55$ and increases
monotonically with time until virialization
(e.g. \citealt{peebles80}).  We can estimate the relevant fraction
with a slight extension of our model: we assume that all gas collapses
spherically, so that a parcel of gas with a given density at $z$ has a
well-defined turnaround redshift $z_{\rm ta}$.  The fraction of gas
with $\delta > \langle \delta_{21} \rangle$ at $z$ is then simply the
fraction of gas that has passed through turnaround shocks by $z_{\rm
ta}$.  Note that our model is only approximate because we do not
follow $(\delta,T_S)$ of individual gas elements; we actually
underestimate the fraction of gas with $T_S>T_K$ because the hottest
shocks, associated with the most massive objects, have smaller
critical densities and hence may have collapsed later on.  The bottom
panel of Figure \ref{fig:21cmemit} compares the fraction of the IGM
gas that is able to emit 21 cm radiation in our shock model (solid
line) with the fraction of gas in minihalos (dashed line).  The peak
at $z \sim 25$ occurs because $\langle \delta_{21} \rangle <
\delta_{\rm ta}$ at higher redshifts, so that all shocked gas was able
to emit.  At $z \ga 15$, the former is much larger than the amount of
mass in minihalos.  Thus, we see explicitly that the true emission
signal will be significantly larger than that calculated just from
minihalos.  Weak shocks occurring before turnaround (as appear to
exist in the simulations of \citealt{gnedin03}) would further amplify
the fluctuation signal.\footnote{Note that the bias amplification for
shocked gas is somewhat smaller than that for minihalos, because it is
associated with smaller density peaks.  However, on the large scales
accessible to the next generation of radio telescopes, the difference
should not be large.}

\section{Discussion}
\label{disc}

In this paper, we have presented a model for the temperature
distribution of IGM gas that takes into account shocks during
structure formation.  We use a simple extension of the Press-Schechter
model that associates shocks with a fixed density contrast, and we
assume that the shock temperature is determined by the infall velocity
at the shock time.  Our model therefore describes moderately overdense
gas in the process of collapsing.  We find that associating shocks
with turnaround adequately describes the temperature distribution of
shocked gas at $z=0$ found by \citet{dave01-whim}.  The agreement with
simulations is not good for weak shocks (which dominate at $z\sim
2$--3), but we are primarily concerned with strong shocks.  We then
used our formalism to consider shocks at higher redshifts.  We found
that large-scale structure (rather than virialization) shocks dominate
the thermal energy content of the IGM for most of the period before
reionization.  As a result, shocks can have several interesting
consequences for structure formation.  These include stripping gas
from $\sim 10\%$ of all minihalos, depositing a substantial fraction
of their energy in cosmic rays, and creating a radiation background
(due to cooling radiation and inverse Compton emission by
accelerated electrons) that extends to high energies.  We have argued
that these shocks can be observed through their effects (both directly
and indirectly) on redshifted 21 cm fluctuations.  The warm,
moderately overdense shocks will emit 21 cm radiation and may be
visible as a fluctuating background.  The fluctuations will be larger
than those from collapsed objects because significantly more gas is
contained in the shocked phase.  Moreover, the X-ray background
produced by shock-accelerated electrons significantly affects the
temperature of gas in the low-density IGM, decreasing (or even
eliminating) any absorption signal.

Our formalism involves several important simplifications.  First, we
assume that infalling gas is shocked only twice: once at turnaround
and once at virialization.  We have also assumed pure spherical
collapse.  This is clearly only an approximation; \citet{keshet03} and
\citet{ryu03} have shown that the distribution of shocks is much more
complicated, with strong shocks bounding filaments and sheets with a
network of weaker shocks surrounding virialized objects.  The
non-spherical nature of collapse is also critical, because it affects
both the fraction of gas that is shocked (through the density
threshold required before significant collapse occurs) and the
strength of the resulting shocks (because the infall velocities also
change).

Another caveat is that our model associates shocks with turnaround.
On the one hand, we seem to overestimate the fraction of gas that is
weakly shocked at moderate redshifts.  We have argued that this is
because gas flows are only moderately supersonic in such shocks.  When
infall velocities are slow, it is plausible that shocks can only form
later in the collapse process.  On the other hand, if gas travels
supersonically during the early stages of infall, asymmetries in the
flow may cause parcels of gas to collide and shock well before
turnaround.  This is particularly important before reionization, when
the IGM gas temperature is $\sim 10 \kel$.  In this regime, supersonic
flow occurs well before turnaround and a much larger fraction of the
IGM may be weakly shocked.  However, the thermal energy of the shocks
is dominated by the largest objects if $z \la 20$ (compare the solid
and dotted curves in Figure \ref{fig:tbar}).  The strong shocks at or
after turnaround therefore probably contain most of the energy.
Furthermore, most of the effects that we consider at high redshifts
(including cooling radiation, particle acceleration, and ram pressure
stripping) depend primarily on the characteristics of the strong
shocks that are able to ionized the cold pre-reionization IGM.  We
therefore expect our conclusions to be robust against uncertainties in
the treatment of weak shocks.  Weak shocks will, however,
substantially increase the 21 cm signal during the early phases of
structure formation (compare our Figure \ref{fig:global} and Figure 2
of \citealt{gnedin03}).

Finally, we have assumed that the \citet{press} formalism provides a
good fit to the mass function.  \citet{sheth} and \citet{jenkins} have
shown that a modified version of the formalism incorporating
ellipsoidal collapse provides a better fit to many simulations (but
see \citealt{jang01}).  We have chosen not to use these modifications
because we require the conditional mass functions to remove mass
attached to collapsed objects.  Conditional functions can be written
analytically in the standard Press-Schechter approach \citep{lacey}.
The Sheth-Tormen approach, however, is based on a ``moving barrier''
for which conditional mass functions cannot be solved analytically,
although they can be fit to simulation data \citep{sheth02}.  For our
purposes, the major difference between the two approaches is that
highly nonlinear objects are more common in the Sheth-Tormen mass
function.  We therefore expect high-temperature shocks to be somewhat
more important than our model predicts.

Obviously, the best way to test our model is with high-resolution
simulations including a careful treatment of shocks.  Such simulations
would follow the non-spherical nature of the collapse directly and can
(at least in principle) capture the shocks at many different stages of
collapse.  However, note that the most significant shocks at high
redshifts (those that ionize the infalling gas) are quite rare.
Large-volume simulations are required in order to include a
representative sample of these rare and highly biased peaks
\citep{barkana03}.  Shock-capturing schemes could then be used in
these simulations to predict more accurately the background radiation
from accelerated particles, the temperature distribution and its
effects 21 cm signal, and the importance of ram pressure stripping.

\acknowledgements S.~R.~F. thanks S.~P. Oh for helpful discussions.
This work was supported in part by NASA grant NAG 5-13292, and by NSF
grants AST-0071019, AST-0204514 (for A.L.).

\end{document}